\documentclass[12pt]{article}

\usepackage{amsfonts,amssymb,amsmath,amscd}

\newcommand{\ra}{\rightarrow}

\newcommand{\RR}{{\mathbb R}}
\newcommand{\CC}{{\mathbb C}}
\newcommand{\tomega}{{\tilde\omega}}
\newcommand{\tg}{{\tilde g}}

\title{Gauge theory, topological strings, and S-duality}

\author{Anton Kapustin\\{\it California Institute of Technology, Pasadena, CA 91125, U.S.A.}}

\begin{document}

\begin{titlepage}

\maketitle

\begin{abstract}

We offer a derivation of the duality between the
topological $U(1)$ gauge theory on a Calabi-Yau 3-fold and the
topological A-model on the same manifold. This duality was
conjectured recently by Iqbal, Nekrasov, Okounkov, and Vafa. We
deduce it from the S-duality of the IIB superstring. We also argue that
the mirror version of this duality relates the topological B-model on
a Calabi-Yau 3-fold and a topological sector of the Type IIA Little
String Theory on the same manifold.

\end{abstract}

\vspace{-5in}

\parbox{\linewidth}
{\small\hfill \shortstack{CALT-68-2490}} 

\vspace{5in}

\end{titlepage}

\section{Introduction}

Recently an interesting connection has been found between
Gromov-Witten invariants of a Calabi-Yau 3-fold $X$ and a
topologically twisted noncommutative $U(1)$ gauge theory on
$X$~\cite{INOV}. The connection is that the generating function of
the Gromov-Witten invariants (in other words, the all-genus
partition function of the A-model for $X$) coincides with the
partition function of the topological $U(1)$ gauge theory.
According to Ref.~\cite{INOV}, the A-model string coupling $\lambda$ is
related to the theta-angle of the gauge theory as follows:
\begin{equation}\label{thetag}
e^{-\lambda}=-e^{i\theta}
\end{equation}

The path integral of the A-model localizes on holomorphic maps
(called holomorphic instantons) from the worldsheet to the target
Calabi-Yau $X$. The weight of a holomorphic instanton depends on
its symplectic area. This theory is ``topological'' in the sense
that for a fixed symplectic form it does not depend on the choice
of a K\"ahler metric, i.e. it is a symplectic invariant. On the
other hand, the path-integral of the gauge theory localizes on
solutions of the Hermitian Yang-Mills equations (HYM) deformed by
terms depending on scalars. In a $U(1)$ gauge theory, these
equations do not have interesting nonsingular solutions,  but if
one makes $X$ noncommutative, nontrivial smooth solutions exist,
which look like four-dimensional instantons wrapping holomorphic
curves in $X$. This gauge theory is also ``topological'', in the
sense that its partition function is a symplectic invariant. This
can be made explicit by writing the action of the gauge theory as
a sum of
\begin{equation} \label{topaction}
\frac{1}{2g_s}\int \omega\wedge F\wedge
F+\frac{i\theta}{6(2\pi)^3}\int F\wedge F\wedge F.
\end{equation}
and BRST-exact terms. Thus the gauge-theory partition function
depends only on $g_s^{-1}\omega$ and $\theta$. As mentioned above,
$\theta$ corresponds to the genus counting parameter of the dual
topological string, while the combination $g_s^{-1}\omega$ is
mapped by the duality to the symplectic form of the A-model. In
Ref.~\cite{INOV} the coincidence of the A-model partition function
and the gauge-theory partition function has been verified in the limit
$\omega\ra\infty$ (i.e. $X$ is replaced by flat space), and more generally
for arbitrary noncompact toric Calabi-Yau
manifolds. It is not clear
how to extend the arguments of Ref.~\cite{INOV} to compact Calabi-Yau manifolds. In
Ref.~\cite{NOP} the conjecture has been reformulated by replacing
solutions of HYM equations on the noncommutative deformation of
$X$ with ideal sheaves on $X$. This reformulation is especially
useful for compact $X$, since for such $X$ it is not completely
clear what is meant by a noncommutative deformation of the gauge
theory.

Even more recently, it was noted that the duality of Ref.~\cite{INOV} should
follow from the S-duality of the Type IIB superstring~\cite{NOV}. In
this note we develop further this idea and propose a simple
physical picture explaining the coincidence of the partition
functions.

\section{Embedding into superstring theory}

In order to embed the 6d topological gauge theory into Type IIB
string theory, one can take a single Euclidean D5 brane wrapped on
the Calabi-Yau manifold $X$. This can be thought of as an
instanton in the remaining four noncompact dimensions. The
theta-angle of the gauge theory on the D5-brane is identified with
the RR 0-form $C_0$ (which is constant for the D5-brane solution).
It is well-known that the low-energy action of the gauge theory on the D5
worldvolume is identical to the action of the topological gauge
theory in $d=6$ obtained by gauge-fixing the action
Eq.~(\ref{topaction})~\cite{6dg1,6dg2}. Thus the partition functions of the two
theories coincide. They are both functions of $\theta$ and
$g_s^{-1}\omega$. To make the gauge theory noncommutative, one can
turn on a flat NS B-field on $X$, as usual~\cite{CDS,DH}.

One may ask about the significance of the partition function of
the D5 brane from the point of view of the effective field theory
in four dimensions. Type IIB superstring compactified on $X$ gives
rise to $N=2$ supergravity which contains $h^{2,1}(X)$ vector
multiplets, $h^{1,1}+1$ hypermultiplets, and the gravity
multiplet. Out of $h^{1,1}+1$ hypermultiplets, the first $h^{1,1}$
come from the K\"ahler moduli of $X$ and their superpartners. The
last one, called the universal hypermultiplet, contains the
dilaton, the dual of the NS B-field (the axion), the RR 0-form
$C_0$, and the dual of the RR 2-form $C_2$. These modes are
constant along $X$, i.e. they come from $h^{0,0}(X)$.

There are two kinds of F-terms in $N=2$ supergravity: the ones
depending only on the vector multiplets and the gravity multiplet
(the latter is described by a chiral Weyl superfield), and the ones depending
only on the hypermultiplets. Since the dilaton sits in a
hypermultiplet, the former F-term cannot receive quantum
corrections and is tree-level exact. In fact it can be computed in
terms of the all-order partition function of the B-model on
$X$~\cite{Aetal,KS}, where the genus expansion of the B-model
corresponds to the expansion of the F-terms in powers of the Weyl
superfield. On the other hand, the hypermultiplet F-terms can
receive corrections, both perturbative and nonperturbative.
Nonperturbative corrections come from Euclidean D-branes or
NS5-branes wrapping $X$. One expects that the contribution of the
D5-instanton to the F-terms is proportional to the D5 partition function.

\section{The derivation of the duality}

The main object of interest will be the partition function of Type IIB string theory in a background with a D5-brane instanton, in the limit when the string coupling $g_s$ goes to zero.
Closed-string
degrees of freedom decouple in this limit and can be regarded as a fixed classical background. 
On the other hand, the effective coupling of the gauge theory on the D5-brane is $\omega/g_s$,
so we will also let $\omega$ go to zero, with the ratio $\omega/g_s$ fixed. The limit
$\omega\ra 0$ is the opposite of the zero-slope limit, where the 6d super-Yang-Mills action
is applicable. However, we will argue below, using S-duality and supersymmetric nonrenormalization
theorems, that higher-derivative terms in the D5-brane action do not contribute to the
partition function, so in the limit we are considering the open-string partition function
coincides with the partition function of the topological gauge theory with an action 
Eq.~(\ref{topaction}). 

We now propose a different way to compute the same partition function. First we perform
S-duality which turns the D5-brane into an NS5-brane wrapped on the same Calabi-Yau.
{}From the point of view of $N=2$ supergravity, the D5-instanton is a singular field configuration
involving the fields of the universal hypermultiplet. After S-duality, the situation is slightly
better: the field configuration corresponding to the NS5-brane instanton is nonsingular, because
of the famous ``throat behavior'' of the metric, but the dilaton grows without bound as one
goes down the throat. Therefore we compactify one direction transverse to the NS5-brane and
perform a further T-duality on it. This transformation turns the NS5-brane into
a Kaluza-Klein monopole. In other words, we end up with Type IIA
string theory on the direct product of $X$ and the Taub-NUT space
$Y$.\footnote{If $C_0\neq 0$, then this statement is precisely true only in the limit $g_s\ra 0$.
For general $g_s$ the Type IIA geometry is a warped product, i.e. one has a fiber bundle with fiber $X$,
such that the metric restricted to any fiber is conformally related to a fixed Calabi-Yau metric, and the conformal factor depends on the base coordinates. We are interested in the limit where $C_0$ is fixed
and $g_s$ goes to zero. In this limit there is no warping (the conformal factor becomes constant).}
After T-duality the string coupling becomes constant (and large). It is also important
to determine the mapping of the RR 0-form. It is easy to check that a constant RR 0-form
is mapped by S and T-dualities to the RR 1-form $C_1$ with an anti-self-dual field-strength.
Explicitly, it is given by
$$
C_1=a V^{-1}(dt+\omega_i dx^i).
$$
Here $V=1+1/r$ is a harmonic function on $\RR^3$ which appears in the
Gibbons-Hawking ansatz for the metric, and $\omega_i$ satisfies
${\rm curl}\ \vec{\omega}={\rm grad}\ V$. This is a unique $U(1)$ instanton on the
Taub-NUT space. The overall factor $a$ parametrizes the asymptotic Wilson loop of $C_1$ in
the $t$-direction. S and T-dualities identify 
$$
a=C_0 \left(\frac{1}{g_s^2}+C_0^2\right)^{-1},
$$
where $C_0$ is the asymptotic value
of the RR 0-form, and $g_s$ is the asymptotic string coupling in the original Type IIB background.

Now note that $C_1$ in Type IIA superstring compactification is
the graviphoton. Thus the composition of S and T dualities maps
the D5 instanton to a smooth four-dimensional Type IIA background
with an anti-self-dual (ASD) metric (the Taub-NUT metric) and an
ASD graviphoton field-strength. The partition function of this background
is the exponential of the quantum effective action of the $N=2$ $d=4$ supergravity.

The full supergravity action is known only to leading order in the derivative expansion.
However, for anti-self-dual metric and the anti-self-dual graviphoton the exact action
can be expressed in terms of the partition function of the topological A-model of the 
Calabi-Yau~\cite{Aetal,KS}, at least if one restricts to terms polynomial in the curvatures.
Namely, for Type IIA theory, only gravitational F-terms of the form
\begin{equation}\label{Fterm}
\int R_-\wedge R_- (\tg_s^2 F_-)^{2g-2} d^4x
\end{equation}
contribute. Here $R_-$ is the ASD part of the Riemann tensor (in the string frame), $F_-$ is
the ASD part of $F_2=dC_1$, and $\tg_s$ is the Type IIA string coupling. If one passes to the
Einstein frame, then this terms becomes independent of $\tg_s$, and since $\tg_s$ sits in
the universal hypermultiplet, which cannot appear in gravitational F-terms, we see that
the coefficient of this term should be $\tg_s$-independent. This result follows from
constraints of supersymmetry, and therefore is valid nonperturbatively.
This is very important for us, since we are interested in the limit $g_s\ra 0$, which
corresponds to the Type IIA string coupling
$$
\tg_s=g_s \left(\frac{1}{g_s^2}+C_0^2\right)
$$
going to infinity.

According to Refs.~\cite{Aetal,KS}, the
coefficient of the F-term Eq.~(\ref{Fterm}) is proportional to $F_g$, the genus-$g$ partition function of the A-model on $X$. For our purposes, the precise coefficient will not be very
important. What is important is that $F_-\sim a$, and therefore the F-term Eq.~(\ref{Fterm})
is proportional to
$$
(\tg_s^2 F_-)^{2g-2}\sim (\tg_s^2 a)^{2g-2}=\left(C_0\left(1+g_s^2 C_0^2\right)\right)^{2g-2}.
$$
Therefore in the limit $g_s\ra 0$ the expansion of the Type IIA free energy in powers of 
$\tg_s^2 F_-$ is the same as the expansion of the D5-brane free energy in powers of $C_0$.

The string partition function $F_g$ is a function of the K\"ahler form $\tomega$ of $X$.
We can relate it to the K\"ahler form before S and T-dualities.
Under S-duality the Calabi-Yau metric gets multiplied by 
$$
\sqrt{\frac{1}{g_s^2}+C_0^2},
$$
while under T-duality in a transverse direction it remains unchanged.
Thus the Type IIA K\"ahler form $\tomega$ is related to the K\"ahler
form $\omega$ in the original set-up (with a D5-brane) by
$$
\tomega=\frac{\omega}{g_s}\sqrt{1+g_s^2 C_0^2}.
$$
In the limit $g_s\ra 0$, $\omega\ra 0$  the Type IIA K\"ahler form
$\tomega$ tends to a definite limit $g_s^{-1}\omega$, which
is precisely the effective coupling of the topological gauge theory.

Now we equate the $C_0$-dependent terms in the free energy of the original Type IIB background
(with a D5-brane wrapped on $X$) and the free energy of the effective field theory obtained
by compactifying Type IIA string on $X$ and turning on $R_-$ and $F_-$. 
Taking into account the identifications of the parameters of the Type IIA and Type IIB backgrounds, 
we come to the following conclusion. Let $\theta_0$
be the value of the $\theta$-angle on the D5-brane for vanishing $C_0$. (It is usually
assumed that $\theta_0=0$, but we will allow for a more general possibility).
Let us expand the free energy of the topological gauge theory on $X$ in
powers of $\theta-\theta_0$, where $\theta_0$ is the value of the
$\theta$-angle for vanishing $C_0$:
$$
\log Z_{D5}=\sum_{g=1}^\infty (\theta-\theta_0)^{2g-2} R_g.
$$
The coefficients $R_g$ are functions of $g_s^{-1}\omega$. Then for $g>1$
we must have
$$
R_g\left(\frac{\omega}{g_s}\right)=b(g) F_g(\tomega),
$$
where $\tomega=g_s^{-1}\omega$, and $b_g$ is some $g$-dependent number.

Note that we only compared $C_0$-dependent parts of the free energies. This is the reason
we restricted the range of $g$ to $g>1$. Thus in essence we are comparing the contributions
to the free energy which are nonperturbative from the gauge theory viewpoint.

It remains to argue that the partition function of the D5-brane can be computed using the topological gauge theory with the action Eq.~(\ref{topaction}). First of all, we expect
that quantum corrections contribute only to the BRST exact terms and can be ignored.
As for stringy tree-level corrections, they are of order $g_s^{-1}$, but have fewer powers
of $\omega$ than the first term in Eq.~(\ref{topaction}) (the net power of $\omega$ can be
negative). In the limit
we are considering, such terms would blow up, and then $R_g$ would not have a well-defined limit.
On the other hand, $F_g(\tomega)$ does have a well-defined limit, which means that 
higher-derivative terms may contribute only to the BRST-exact terms in the gauge theory action.

One can determine the numerical coefficients $b_g$, as well as $\theta_0$, by computing
both $R_g$ and $F_g$ in some particular case and comparing them. In
Ref.~\cite{INOV} this has been done for $X=\CC^3$, and it was found that the topological
string coupling and the $\theta$-angle are related by Eq.~(\ref{thetag}). This means that
$$
R_g=(-1)^{g-1} F_g,\quad \theta_0=\pi.
$$
This seems to suggest that for vanishing $C_0$ the
theta-angle of the D5-brane theory is $\pi$, rather than $0$. Note
that $\theta=\pi$ does not break parity-invariance of the
worldvolume theory, so there is no obvious contradiction here.

\section{Discussion}

We showed that S-duality of Type IIB string theory implies the
conjecture of Ref.~\cite{INOV}. 
An interesting question is the role of noncommutativity of
the D5 worldvolume. In the gauge theory computation, it serves as
a regulator which gives $U(1)$ instantons (i.e. D-strings) finite
size. From the string theory viewpoint, noncommutativity comes
from the NS B-field~\cite{CDS,DH}. From the viewpoint of the
worldvolume theory of the D-strings, it is a Fayet-Iliopoulos term
which lifts the Coulomb branch on the D-string worldvolume theory.
Ordinarily, this Coulomb branch describes the motion of D-strings
in the directions transverse to the D5-brane. Thus turning on
noncommutativity makes D-strings ``stick'' to the D5-brane. Upon
S-duality, the NS B-field turns into a (flat) RR 2-form $C_2$.
Thus one expects that a flat $C_2$ makes the F-strings ``stick''
to the Type IIB NS5-brane. This may seem strange, since flat RR
fields usually do not have any effect on F-strings. However, this
is only true for vanishing NS field $H_3$. In general, $C_2\wedge
H_3$ serves as a source for the RR 5-form flux $F_5=dC_4$, which
can affect F-strings. It would be interesting to understand this
effect in detail. Here we only make the following simple
observation: the fact that flat B-field gives D-strings a finite
size can be explained after S-duality in terms of the Myers
effect~\cite{myers,giant}. Namely, the RR 5-form flux makes F-strings
expand into D3-branes stuck to the D5 worldvolume. It appears that
this effect is responsible for the ``sticking'' of the F-strings to the
NS5 worldvolume.

Another interesting topic is the mirror version of this
duality. This issue has been raised and discussed in Ref.~\cite{NOV}.
We would like to point out that the answer to this question is essentially
contained in a paper by Dijkgraaf, Verlinde, and Vonk~\cite{DVV}. 
The mirror statement is that the all-order
B-model partition function on $X$ computes the partition function
of the Type IIA NS5-brane wrapped on $X$. The topological string
coupling is dual to the expectation value of the RR 3-form on $X$
(which is proportional to the holomorphic 3-form on $X$). The
derivation of this duality in Ref.~\cite{DVV} is almost
the same as above: one starts with a Type IIA NS5-brane on $X$, performs
T-duality, and ends up with a IIB Kaluza-Klein monopole wrapped on
$X$. Then one identifies the gravitational F-terms in Type IIB string evaluated
on a Kaluza-Klein monopole with the free energy of the Type IIA NS5-brane.
One difference with the above derivation is that one does not have to appeal to S-duality.
Another difference is that we have no independent way of
computing the quantum partition function of the Type IIA
NS5-brane. Type IIA NS5-brane in the limit $\tg_s\ra 0$
is described by Little String Theory, which loosely speaking describes
self-dual strings. The results of Ref.~\cite{DVV} show that the
topological sector of the Little String Theory is equivalent to the
topological string theory of type B. 

Nekrasov, Ooguri, and Vafa proposed~\cite{NOV}
that the B-model partition function of $X$ is dual to the partition function
which ``counts'' special Lagrangian submanifolds in $X$. According
to this conjecture, the topological string coupling is dual to the expectation 
value of the RR 3-form. Thus this proposal seems to be closely related to the
results of Ref.~\cite{DVV} relating Type IIA LST on $X$ and the B-model on $X$.
One may speculate that the Type IIA LST admits some sort of BPS membranes, which can be
thought of as Euclidean D2-branes stuck to the NS5 worldvolume, and that
the path-integral of the LST localizes on Euclidean membranes wrapping special Lagrangian
3-cycles in $X$. By analogy with the case of D5-branes, one expects that such BPS membranes
exist as nonsingular solutions after one regularizes the LST, by turning on
a suitable flat RR form (the analogue of the B-field). Since $X$ is simply connected,
the only candidate RR form is the 3-form $C_3$. 
It would be very interesting to prove or disprove these conjectures.

\section*{Acknowledgments}

I am grateful to Andrei Mikhailov and Hiroshi Ooguri for
discussions. I also would like to thank Jaume Gomis for pointing out an inaccuracy
in the first version of the paper. This work was supported in part by the DOE grant DE-FG03-92-ER40701.

\end{document}